\documentclass[11pt]{article}
\usepackage{amssymb,amsmath,amsthm,graphicx,color,setspace,mathrsfs, mathtools, newpxtext, newpxmath, 
	multirow, subcaption, natbib, bbm, threeparttable, caption, xcolor, array, booktabs, fancyhdr}
\usepackage[hidelinks]{hyperref}
\usepackage[capposition=top]{floatrow}
\usepackage[margin=1in]{geometry}
\usepackage[title]{appendix}

\newtheorem{assumption}{Assumption}

\begin{document}
	
	\onehalfspacing
\title{The Identity Fragmentation Bias}
\author{Tesary Lin\thanks{Boston University Questrom School of Business;
		\href{mailto:tesary@bu.edu}{tesary@bu.edu;}}
	\and
	Sanjog Misra\thanks{The University of Chicago Booth School of Business;
		\href{mailto:sanjog.misra@chicagobooth.edu}{sanjog.misra@chicagobooth.edu}. 
		The authors thank Dean Eckles, Max Farrell, Garrett Johnson, and Harikesh Nair for helpful comments and suggestions. }
}
\date{\today}
\maketitle
\thispagestyle{empty}

\begin{abstract}
	Consumers interact with firms across multiple devices, browsers, and machines; these interactions are often recorded with different identifiers for the same consumer. The failure to correctly match different identities leads to a fragmented view of exposures and behaviors. 
	This paper studies the \textit{identity fragmentation bias}, referring to the estimation bias resulted from using fragmented data. 
	Using a formal framework, we decompose the contributing factors of the estimation bias caused by data fragmentation and discuss the direction of bias. Contrary to conventional wisdom, this bias cannot be signed or bounded under standard assumptions. Instead, upward biases and sign reversals can occur even in experimental settings. We then compare several corrective measures, and discuss their respective advantages and caveats. 		
	
	\vspace{0.25cm}
	
\end{abstract}

	\newpage
	\setcounter{page}{1}

	\section{Introduction}

Consumers' digital footprints are becoming increasingly fragmented. A typical consumer uses multiple devices and navigates across websites throughout an online purchase journey. Companies and websites typically track consumers via cookies, which are text files generated to identify a \textit{user agent} (a browser-device combination) when consumers first visit a website. However, cookies are browser, device, and site specific. As such, a consumer is often associated with multiple, mutually disconnected cookie identifiers across websites and browsers, with each one capturing only a fraction of their ad exposures and behaviors. We call this fracture of identifiers and records \textbf{identity fragmentation}. 

Websites often deploy several workarounds for the identity fragmentation problem. One solution is using a device ID to identify users. For example, mobile apps frequently obtain IDFA (advertiser ID for Apple devices) and Android's Advertiser ID for tracking. Using a device-level identifier alleviates but does not completely avoid fragmentation, since a user can often navigate through multiple devices within the same journey. For example, Facebook reports that 32\% of  users who show interest in their mobile ads convert on their desktop.\footnote{
	https://www.facebook.com/business/news/cross-device-measurement
} Alternatively, advertising platforms sometimes perform cookie-syncing, which matches several fragmented records across publishers and ad-tech partners. Cookie syncing can match only up to 60\% of fragmented records\footnote{
	https://www.adexchanger.com/data-driven-thinking/ad-tech-needs-a-shared-id-solution-asap
}, while slowing down the ad loading by 19 seconds on average\footnote{
	https://headerbidding.co/cookie-syncing/\#privacy
}. Despite the short-comings of device IDs and cookie syncing, they are the preferred workarounds for most websites. Compared to alternatives such as a login wall, these two approaches can often be implemented without users' explicit consent, thus minimizes friction and deterrence of potential customers. 

Unfortunately, the advantages of device IDs and synced cookies also make them the target of industry self-regulation that responds to consumers' demand for privacy. Starting with iOS 14, Apple requires that mobile apps seek users' opt-in consent before collecting their IDFA. On the web end, Chrome ultimately follows Firefox and Safari, announcing that they will stop supporting 3rd-party cookies starting in 2021, also citing privacy concerns. Since all cookie syncing relies on third-party cookies, Chrome's move essentially makes cookie syncing obsolete. As the world becomes more and more attuned to user privacy, identity fragmentation is here to stay.

This paper analytically characterizes how fragmented identifiers bias model estimates and inference, which we call the \textbf{identity fragmentation bias}. In its simplest form, the problem manifests as an inability to size the user group. More importantly, causal associations across cookies and devices cannot be reliably measured without a unique identifier. Consider a consumer who sees an ad on her desktop and later buys the advertised product on her phone. If these interactions are taken as independent observations from different users, the causal link between the ad exposure and the purchase is broken. Researchers and firms who use fragmented data may thus estimate the advertising effect with bias.

We decompose the bias into three components in the context of a linear model. The first component arises because variations in the outcome variable are split. As a result, each fragmented identifier only captures a fraction of the outcome variation, leading to an attenuation bias. The second component emerges from the fracturing of covariates. This fracture prevents the variation of covariates in the mobile record from being associated with outcomes in the laptop record and vice versa, creating an omitted variable bias. The third component is caused by a spurious correlation between fragmented outcomes and covariates. This can occur with a device-level activity bias or a cross-device substitution triggered by ad exposure. The latter two bias components have arbitrary signs and magnitudes. As a result, the direction of the identity fragmentation bias is undetermined without further assumptions. Moreover, this bias does not converge to zero in the limit. We show that these conclusions hold regardless of whether the model assumes the estimate of interest is homogeneous or device-specific. 

We discuss special cases where the bias can be completely characterized. In particular, we show that when an experiment places equal treatment probabilities (for binary treatments) or intensities (continuous treatments) across sites and devices, the raw estimate is attenuated by a factor equal to the number of fragmented records per user. In this case, we can obtain an unbiased estimate for the true parameter by multiplying the raw estimate with the number of fragments. This finding is closely related to \citet{coey2016people}, who propose a debiasing estimator for cookie-level estimates under similar conditions.

We show the robustness of this result by examining a number of model perturbations. First, we compare models where the parameters of interest are common across fragments and models where parameters of interest are fragment-specific. Then we examine a setting with an arbitrary number of fragments. We have also analyzed cases where data contain a mixture of fragmented and complete identities. %, and cases where identities are probabilistically matched. 
In all cases, the identity fragmentation bias persists and is unsigned without further assumptions.

Guided by the analytic form, we compare several solutions for the identity fragmentation bias in the general setting. Existing solutions can be categorized into two buckets. The first is identity linking (with cookie syncing being one example), which works by identifying and matching fragmented records that belong to the same user. This approach is widely adopted in the industry. However, even the best identity linking methods only stitches up part of the fragmented records. We show that partially matched data often complicates the bias pattern and prevents estimator bounding and debiasing. The second approach, proposed by \citet{coey2016people}, adjusts post-experiment estimates based on a set of symmetry assumptions. When the assumptions hold, this approach can effectively debias the average treatment effect estimates without the need to stitch up the raw data.

We then propose a third approach, the \textbf{stratified aggregation}. Instead of constructing user-level records, this approach works by constructing user group-level records (aggregation) and refining the group using covariates (stratification). The estimate after stratified aggregation is unbiased, but with a higher variance. By relaxing the need for constructing user-level data, stratified aggregation can achieve a wider coverage, thereby avoiding the complications caused by identity linking. It also works without additional assumptions or an experiment. This approach works better when the covariates for constructing the stratified groups are high-dimensional, and when the dataset is large enough to account for the increase in variance.

To illustrate the bias and evaluate stratified aggregation, we conduct an empirical exercise using data from an online retailer. In this setting, consumers interact with the seller across mobile, desktop, and tablet devices; their identities are made known to us by an identity matching vendor. We find that when the data is fragmented at the device level, the ad effect estimates are biased upwards. This upward bias is caused by a dominant device-level activity bias. 
Using constructed demographic variables, we show that stratified aggregation can remove the bias, but comes at the cost of precision.

Our paper contributes to several strands of literature. The first one examines the role of customer identifiers for delivering accurate insights, including \citet{rossi1996, aziz2015cookie, trusov2016crumbs}, and \citet{miller2017}. While these papers demonstrate the value of unfragmented data for profiling or prediction, we focus on the impact of fragmented data on parameter estimates. 
More recent papers have documented the economic values of cookie-level data, including \citet{goldfarb2011v2, marotta2019}; and \citet{johnson2020}. We complement their analysis by describing how the inability to persistently track users will impact measurement and model estimation.

Second is the literature that recognizes problems from using cookie-level in lieu of individual-level data, including \citet{chatterjee2003modeling,manchanda2006effect, rutz2011,bleier2015personalized,hoban2015effects}; and \citet{blake2016returns}. While these papers articulate the possibility of problems, we formally characterize the bias that emerges from identity fragmentation. The analytic form allows us to evaluate and compare different solutions to tackle the fragmentation bias.

Our paper is closely related to \citet{coey2016people}, who also examine the analytic form of estimation bias caused by identity fragmentation. They show that the use of cookie-level data leads to an attenuation bias in a setting free from confounds such as activity bias \citep{lewis2015measuring,johnson2017ghost} and cross-channel (device or cookie) substitution \citep{goldfarb2011}. This setting allows them to perform post-experiment adjustment to debias cookie-level estimates. We take a step forward by accounting for activity bias and cross-channel substitution when characterizing the bias. We also relax their assumption that cookie uses are ex-ante symmetric, so that our framework can be generalized to discuss other cross-channel settings.

Several authors have proposed approaches to alleviate the data fragmentation problem. These approaches include data linking (see Abramitzky et al. \citeyear{abramitzky2019} and Bailey et al. \citeyear{bailey2017} for a review), experiment-based adjustments \citep{coey2016people,koehler2016}, missing data imputation \citep{novak2015bayesian}, and aggregation \citep{rutz2011,blake2018}. Our paper takes another step forward by comparing existing solutions and proposing a method to improve the last approach. In particular, we propose refining the aggregation using a combination of variables associated with the fragmented identities, in a way similar to what the industry calls “fingerprinting” but less restrictive.\footnote{This paper is also related to the broad econometrics literature on overcoming unknown network inteference, such as Manresa (2016) and Savje et al. (2019).}

We organize the rest of the paper as follows. Section 2 illustrates the fragmentation problem using an example. Section 3 formally characterizes the bias, followed by a discussion on bias in experiment settings and generalizations. Section 4 compares several solutions, and Section 5 concludes. 
%. We conclude with a discussion and ideas for future research.

\section{An Illustrative Example}\label{example}

	Consider a setting where consumers each has two devices, a mobile (M) and a desktop (D). Consumers are exposed to ads, and the researcher seeks to estimate the impact of ads on purchases. Suppose the advertising effect does not hinge on which device is used for ad delivery. Table \ref{tab:example}(a) presents two consumers who have been exposed to $2$ and $4$ ads, respectively. Since both consumers purchase equal amounts, advertising has no discernible effect $\left(\beta=0\right)$. . 

\begin{table}[h!]
	\centering
	\caption{Estimating Advertising Effects with Fragmented Data: An Example}
	\label{tab:example}
	\centering
	\subcaption{True Data }
	
	\begin{tabular}{p{2cm}|p{2cm}|p{2cm}|p{2cm}|p{2.5cm}} 
		\hline 
		\multicolumn{1}{l|}{Identity} & Device & Purchase  & Ads  & Actual Effect \\ 
		\hline \multirow{2}{*}{1}    & D      & 	\multirow{2}{*}{1} & \multirow{2}{*}{2} & \multirow{4}{*}{$\beta = 0$}   \\ 
		\cline{2-2}     & M      &                    &                    &                                      \\ 
		\cline{1-4} \multirow{2}{*}{2}          & D      & \multirow{2}{*}{1} & \multirow{2}{*}{4} &  \\ 
		\cline{2-2}    & M      &                    &                    &            \\ 
		\hline 
	\end{tabular}
	
	\vspace{0.25cm}
	
	\subcaption{Fragmented Data}
	
	\begin{tabular}{p{2cm}|p{2cm}|p{2cm}|p{2cm}|p{2.5cm}} 
		\hline 
		\multicolumn{1}{l|}{Observed ID} & Device & Purchase & Ads & Est. Effect \\ 
		\hline 
		1(a)                             & D      & 1        & 2        & \multirow{4}{*}{$\beta = 0.4$} \\ 
		\cline{1-4} 1(b)                           & M      & 0        & 0        &                                                \\ 
		\cline{1-4} 2(a)                            & D      & 1        & 3        &                                                 \\ 
		\cline{1-4} 2(b)                           & M      & 0        & 1        &                                                \\ 
		\hline 
	\end{tabular} 
	
	\vspace{0.25cm}
	
	\subcaption{Fragmented Data}
	
	\begin{tabular}{p{2cm}|p{2cm}|p{2cm}|p{2cm}|p{2.5cm}}
		\hline 
		\multicolumn{1}{l|}{Observed ID} & Device  & Purchase  & Ads  & Est. Effect \\ 
		\hline 1(a)  & D  & 1  & 0 & \multirow{4}{*}{$\beta=-0.4$} \\
		\cline{1-4} 	1(b)  & M  & 0 & 2 & \\
		\cline{1-4}  2(a)  & D  & 1 & 1  & \\
		\cline{1-4} 2(b)  & M  & 0  & 3 & \\
		\hline 
	\end{tabular}
	
\end{table}

Now consider the data where identities are fragmented. Since the researcher can no longer associate devices to consumers, she mistakenly assumes that there are four distinct observations. Table 1(b) shows that if consumers see ads on devices that they then purchase on, a spurious positive correlation occurs, creating an upward bias in the estimated advertising effect. In this example, the parameter estimate is $\beta=0.4$. In other cases, the bias can go in the opposite direction. Suppose instead that consumers see ads more often on their phones but make purchases predominantly on their desktops. Then the estimated advertising effect can be negative, as in panel (c) of Table 1.

This stylized example shows that identity fragmentation can lead to significant biases, which cannot be signed without additional information. In the next section, we formally characterize the identity fragmentation bias and decompose it into three components. We then discuss which feature of fragmented data creates each bias component, and use the finding to guide our discussion on debiasing approaches.

\section{Characterizing the Identity Fragmentation Bias}
	\label{formal}

We use the term ``fragment'' to describe any sub-identity that a consumer may have. Fragments can arise across \emph{cookie}s, \emph{device}s, \emph{channels}, or any other measurement unit across which identities are possibly split. To make the illustration concrete, we focus on the earlier example, where the research goal is estimating the effect of advertising on purchases and where data are fragmented across devices. The analysis and results are not constrained to this example. 

To facilitate exposition, we consider estimates under the standard linear model. \footnote{
	The sources of bias we present here persists in nonlinear models, though the exact expression of the bias may change. 
} In Section \ref{main-model}, we assume that each consumer has two devices and that the researcher cannot separate observation units by device types. As such, the researcher does not distinguish ad effect across device types when analyzing the data. We relax this assumption in Section \ref{extension} and onwards.

\subsection{Analysis Framework}\label{main-model}

The researcher estimates a common effect of advertising exposures on purchases:
\begin{equation}
	y = \alpha + x' \beta + \epsilon.
	\label{dgp}
\end{equation}
Here, $y$ denotes the dollar spent per user; $x$ represents a $K$-dimensional vector of covariates including advertising exposure, and $\epsilon$ is the error component. The parameter of interest is $\beta.$  
Let $y_j$, $x_j$ denote the corresponding variables on device $j \in \{1,2\}$. By construction, $y=y_{1}+y_{2}$ and $x=x_{1}+x_{2}$, representing the aggregate spend and exposure levels for a given user. As usual, we also assume $E[\epsilon | x_1,x_2] = 0$.

The un-fragmented (true) data consists of $N$ observations, reflecting $N$ unique consumers whose observations are identically and independently distributed. Let $Y=[y_{\left(1\right)},...,y_{\left(N\right)}]'$, $X_{j}=[x'_{\left(1\right)j},...,x_{\left(N\right)j}']$, and define $X=[\eta\quad X_{1}+X_{2}]$ where $\eta$ is a length-$N$ vector of ones. If consumer-level identities were observed, a researcher could obtain an estimate of $\beta$ by regressing $Y$ on $X$. When the data is \textit{fragmented}, however, the researcher observes 
$$
\tilde{Y}\equiv\left[\begin{array}{c}
	Y_{1}\\
	Y_{2}
\end{array}\right],\quad\tilde{X}\equiv\begin{bmatrix}\eta & X_{1}\\
	\eta & X_{2}
\end{bmatrix}.
$$
Without additional information, the researcher treats each device as a different user, assumes that the number of observations is $2N$, and estimates $\beta$ by regressing $\tilde{Y}$ on $\tilde{X}$.

An important aspect of identity fragmentation is how purchases occur across devices. We allow consumers to have idiosyncratic preferences over how they use devices for purchases, and capture these preference using a binary variable $s_{\left(i\right)}$. In particular, $s_{\left(i\right)}=1$ implies that consumer $i$ completed his purchases on device 1; $s_{\left(i\right)}=0$ means his purchase was made on device 2.\footnote{
	Making $s_{\left(i\right)}$ a continuous variable $\in [0, 1]$ does not change our results. The only reason that we assume $s_{\left(i\right)}=1$ to be binary is that it is unusual for consumers to split their purchases across devices within a shopping journey.
} We define \textbf{$\mathbf{s}$} to be an $N\times N$ diagonal matrix where the $i_{th}$ diagonal is $s_{\left(i\right)}$, and stack these matrices to form
$S=\begin{bmatrix}\mathbf{s}\\
	I-\mathbf{s}
\end{bmatrix}$. 
We can then express the relationship between the fragmented and un-fragmented purchases as 
\begin{equation}
	\tilde{Y}=SY.
\end{equation}
We define the expectation of device usage conditional on covariates as $\Lambda_{x}=\mathbb{E}[\mathbf{s}|(X_{1},X_{2})]$. Further, we define $W\equiv[I_{N\times N}\quad I_{N\times N}]$, and $\Omega_{(K+1)\times(K+1)}=diag(1/2,1_K)$. With this notation, we characterize the relation between the unfragmented covariates $\left(X\right)$ and its fragmented version $\left(\widetilde{X}\right)$ as 
\begin{equation}
	X=W\widetilde{X}\Omega.
\end{equation}
Let $\theta \equiv 
\begin{bmatrix}
	\alpha\\
	\beta
\end{bmatrix}$. The regression estimator using fragmented data can now be written as 

\begin{align}
	\widehat{\theta} & =(\widetilde{X}'\widetilde{X})^{-1}(\widetilde{X}'\widetilde{Y})\\
	& =(\widetilde{X}'\widetilde{X})^{-1}(\widetilde{X}'SY)\\
	& =(\widetilde{X}'\widetilde{X})^{-1}[\widetilde{X}'S(W\widetilde{X}'\Omega\theta+\epsilon)]
\end{align}
We introduce the following assumptions to exclude other conventional forms of bias, so that the remaining bias is induced \textit{solely} by identity fragmentation:
\begin{assumption} 
	$s\perp\epsilon|(x_{1},x_{2})$; 
\end{assumption}
\begin{assumption} 
	Standard OLS assumptions: exogeneity $(\mathbb{E}[\epsilon|X]=0)$, no perfect collinearity $(\mathbb{E}[XX']>0)$, i.i.d. and homoskedastic errors $\epsilon$. 
\end{assumption} 
Under these assumptions, we can represent the conditional expectation of $\theta$ as
\begin{equation}
	\mathbb{E}[\widehat{\theta}|X_{1},X_{2}]=\Omega\theta+(\widetilde{X}'\widetilde{X})^{-1}
	\begin{bmatrix}0\\
		(X_{1}-X_{2})'[\Lambda_{x}-\frac{1}{2}I]\eta\alpha+[-X_{1}'(I-\Lambda_{x})X_{1}+X_{1}'X_{2}-X_{2}'\Lambda_{x}X_{2}]\beta
	\end{bmatrix}.
\end{equation}

Our main interest is $\widehat{\beta}$, the estimate for the slope parameter(s). Let $\vartheta$ be the bottom-right sub-matrix of $(\widetilde{X}'\widetilde{X})^{-1}$,
then the conditional mean of estimation bias for\textbf{ $\beta$} is (see proof in Appendix \ref{proof-common}):
\begin{align}
	\label{bias-expression}
	\mathbb{E}[\widehat{\beta}|X_{1},X_{2}]-\beta & =\vartheta\left(\underbrace{(X_{1}-X_{2})'[\Lambda_{x}-\frac{1}{2}I]\eta\alpha}_{\Delta_{3}}+\underbrace{[X_{1}'X_{2}]\beta}_{\Delta_{2}} \underbrace{-[X_{1}'(I-\Lambda_{x})X_{1}+X_{2}'\Lambda_{x}X_{2}]\beta}_{\Delta_{1}}\right)\\
	& =\vartheta\left(\Delta_{3}+\Delta_{2}+\Delta_{1}\right).\nonumber 
\end{align}

\subsection{The Bias Components}

The multiplier $\vartheta$ in Equation (\ref{bias-expression}) is always positive definite. Therefore, the sign of bias in $\widehat{\beta}$ is solely determined by the components $\Delta_{1}$, $\Delta_{2}$, and $\Delta_{3}$. Each component represents a specific form of distortion emerging from the split identities: fragmentation of the outcome variable $\left(\Delta_{1}\right)$, fragmentation of the exposure variables $\left(\Delta_{2}\right)$, and interaction of the fragmented outcome and exposures $\left(\Delta_{3}\right)$.

\subsubsection*{$\Delta_{1}:$ Purchase Fragmentation}

The first bias component results from splits of the outcome variable $Y$. Each row in the fragmented data only captures a fraction of variation in the original outcome. This type of measurement error manifests itself via the terms $-X_{1}'(I-\Lambda_{x})X_{1}$ and $-X_{2}'\Lambda_{x}X_{2}$, creating an attenuation bias. To see this mathematically, note that $- I < \vartheta \Delta_1 < 0$. Intuitively, since the number of observations is artificially doubled while the total variation in outcome is constant, there is a direct attenuation on the measured outcome variation, and subsequently, the estimate. In our advertising example, the fact that the average customer spend is lower simply because we think there are $2N$ observations makes advertising seem less effective than it actually is. This component of the fragmentation bias always exists, except in the trivial case when $\beta=0$. 

\subsubsection*{$\Delta_{2}:$ Exposure Fragmentation}

The second bias component comes from the fragmentation of data on the exposure side. The fragmentation of identities prevents the model from establishing the association between $X$'s and $Y$'s \emph{across} fragments. To see this, note that the true covariate is $X=X_{1}+X_{2}.$ However, in the fragmented data, some rows have $\widetilde{X}=X_{1}$ and consequently, the variation pertaining to $X_{2}$ is omitted; in other rows, the opposite occurs. This omission is represented by the interaction term $X_{1}'X_{2}$ in $\Delta_{2}$, which is often seen in the characterization of omitted variable bias (see, e.g., Greene \citeyear{greene2003} and Wooldridge \citeyear{wooldridge2009}).

The direction and magnitude of this component depend on $X_{1}'X_{2}$. When covariates are normalized to mean-zero before entering the model, $X_{1}'X_{2}$ is the correlation of ad exposures across devices. Thus, a stronger, positive (negative) correlation between cross-device ad exposures leads to a positive (negative), larger bias component $\Delta_2$, the same as what occurs for omitted variable bias. More generally, we see that covariate normalization can change the magnitude of this bias term. In the special case where ad exposures are independent across devices, normalizing covariates to mean zero can eliminate this bias component.

\subsubsection*{$\Delta_{3}:$ Spurious Covariance}

This term captures a \textit{spurious} correlation between $\tilde{Y}$ and $\tilde{X}$. Recall that $\tilde{Y}=SY$ and that 
$E[S|(X_{1},X_{2})]=\left[\begin{array}{c}
	\Lambda_{x}\\
	I-\Lambda_{x}
\end{array}\right]$. 
The main element in $\Delta_{3}$, the product $(X_{1}-X_{2})' (\Lambda_{x}-\frac{1}{2}I)$, is the estimate of $\text{\textbf{cov}}(\widetilde{X},S)$. This is the covariance between advertising exposure and device usage preferences. This covariance can be nonzero when there is a device-level activity bias. For example, if a consumer is more likely to see ads and buy products on the same device, then $\mathbf{\text{\textbf{cov}}}(\widetilde{X},S)>0$, creating an upward distortion of the estimator. If the ads and purchases are made systematically on different devices, the resultant distortion goes in the opposite direction. The device-level activity bias can impact the estimate even when the device usage preference is independent of exposure levels. That is, even when $\Lambda$ does not depend on $(X_{1},X_{2})$, $\text{\textbf{cov}}(\widetilde{X},S)\neq0$ as long as $X_{1}\neq X_{2}$ and $\Lambda\neq\frac{1}{2}I$. As long as there are differential device usage preferences and differential exposure levels, this bias term will persist.

Spurious correlation can also occur when consumers substitute their purchases across devices in response to advertising. For example, suppose in response to a promotion on his phone, a consumer increases the purchases via the phone, but at the same time decrease purchases on his desktop by the same amount. The advertising is ineffective overall. However, the researcher overestimates the advertising effect, because she only observes the positive ad-purchase correlation on the phone and interprets the decrease of consumption on the second desktop as noise. From an econometric standpoint, this bias is caused by the fact that the fragmented measures for purchases, $SY$ and $(1-S)Y$, have measurement errors correlated with $X$. 

Activity bias and cross-device substitution represent different consumer behaviors and should be tackled differently. Activity bias results from consumers' device usage preferences itself, while cross-device substitution effects arise due to a shift in device usage caused by ad exposures. As we will see later, activity bias can be removed by unstacking the fragmented data (i.e., estimating separate models by device). By contrast, cross-device substitution effects are not eliminated even when data are unstacked.

We note that $\Delta_{1} + \Delta_{2}$ is bounded by $[-\beta, \beta]$. Therefore, in special cases where the baseline outcome $\alpha$ is close to zero, the bias caused by spurious correlation is negligible, and $\widehat{\beta} \in [0, 2 \beta]$ (or $[2 \beta, 0]$ when $\widehat{\beta}$ is negative). In other words, in this case, the estimate has its sign consistent with the true effect, and also gives us an effect size lower bound $|\beta| > |\widehat{\beta}/2|$. A close-to-zero baseline outcome is more likely to occur for new product sales. On the other hand, in most digital ad effect studies, $\alpha >> \beta$, and the bias caused by spurious correlation should be taken into account.

\subsection{How Does Experiment Help?}
\label{randomization}	

In certain conditions, the bias exhibits only attenuation bias with a simple structure. We characterize the condition below:

\textit{The \textbf{Symmetric and independent exposures (SIE)} condition is satisfied when both of the following statements hold:
	\begin{enumerate}
		\item $E[x_1] = E[x_2]$, $Var[x_1^2] = Var[x_2^2]$; $x_1 \perp x_2$;
		\item $E[\mathbf{s}|X_{1},X_{2}]=\Lambda$ does not depend on $(X_{1},X_{2})$.
	\end{enumerate}
}
Under \textit{SIE}, $E[\widehat{\beta}]=\frac{\beta}{2}$ as long as covariates are normalized to mean-zero before entering the model. Put differently: if the researcher can guarantee that the exposure across fragments is symmetric in mean and variance, and that the device preferences do not depend on exposures, then the estimator is attenuated by a factor exactly equal to the number of fragments. In \citet{coey2016people}, attenuation bias occurs when $x_{1},x_{2}$ are i.i.d. distributed and a user purchases products using each cookie with equal probability: this is a special case of \textit{SIE}.

We note that both conditions in \textit{SIE} are imperative to guarantee an attenuation bias. In particular, suppose we have an experiment, so that $X_1 \perp X_2$, but different fragment types receive different intensities of the treatment. The direction of fragmentation bias will still be arbitrary in this case. Consider the following numerical example:
$$
E[ x_{1}]=3, E[x_{2}]=1, E[x_{1}^{2}]=10, E[x_{2}^{2}]=2; x_1 \perp x_2.
$$
Assume that device preferences are independent of exposure so that
$\mathbb{E}\left(s|x\right)=\lambda_{x}=\lambda$; and that $\beta>0$, $\alpha=0.$ Under this setting, the fragmentation bias is $$
\mathbb{E}[\beta|X_{1},X_{2}]-\beta=\left(2\lambda-\frac{7}{4}\right)\beta.
$$
It is straightforward to see that the bias varies with device preferences $\lambda$:
\begin{align*}
	\lambda & \in(7/8,1]\implies\widehat{\beta}>\beta;\\
	\lambda & \in(3/8,7/8)\implies0<\widehat{\beta}<\beta;\\
	\lambda & \in[0,3/8)\implies\widehat{\beta}<0.
\end{align*}
In other words, randomization by itself does not provide additional guarantees regarding the sign of the fragmentation bias.

An experiment that completely randomizes treatments across device types can help ensure that the first condition in \textit{SIE} is satisfied. On the other hand, condition 2 is an assumption on consumer behavior, which is more likely to be satisfied in certain analysis paradigms than others. We return to this point in Section \ref{solution} where we discuss experiment-based debiasing solutions.

\subsection{Generalizations}\label{extension}
\subsubsection*{Device-Specific Effects}
When device types are well-defined (e.g.,\ consumers use phones and computers; in contrast, cookies associated with the same user may all belong to the same type), a researcher may estimate device-specific advertising effects. They may choose one of the two approaches below, which lead to different forms of biases. The first specifies device-specific ad variables while still pooling data across devices. The model is 
\begin{equation}
	y = \alpha + x_1 ' \beta_1 + x_2 ' \beta_2 + \epsilon.
\end{equation}
The stacked data is now $
\tilde{X} \equiv 
\left[ \begin{array}{ccc}
	\eta & X_1 & \emptyset \\
	\eta&  \emptyset & X_2 
\end{array} \right].
$ 
Let $\widehat{\beta} \equiv [\widehat{\beta}_1' \quad \widehat{\beta}_2']'$. Then the bias can be expressed as the following (see proof in Appendix \ref{proof-specific}):

\begin{equation}
	E\left[\widehat{\beta}|X_1, X_2 \right]  - \beta = \vartheta 
	\begin{bmatrix}
		\underbrace{X_1' (\Lambda_x - \frac{1}{2}I) \eta \alpha}_{\Delta_3} \underbrace{- X_1' (I - \Lambda_x) X_1 \beta_1}_{\Delta_{1}} + \underbrace{X_2' \Lambda_x X_1 \beta_2 }_{\Delta_{2}}\\
		\underbrace{X_2' (\frac{1}{2}I - \Lambda_x) \eta \alpha}_{\Delta_3} +  \underbrace{X_2' (I - \Lambda_x) X_1 \beta_1}_{\Delta_{2}} \underbrace{- X_2' \Lambda_x X_2 \beta_2}_{\Delta_{1}}
	\end{bmatrix}
	\label{bias_b2}
\end{equation}
As before, the estimation bias contains attenuation effect ($\Delta_{1}$), omitted variable bias ($\Delta_{2}$), and spurious correlation ($\Delta_{3}$). However, in contrast to the common-effect model, here consumers' substitution between fragments influences not only the spurious correlation but also the omitted variable bias, as is reflected by the interaction between $\Lambda_x$ and $X_1$, $X_2$.

The second approach splits data by device type and estimates separate ad effect models. The fragmentation bias in this case is 
$$
E\left[\widehat{\beta}|X_1, X_2 \right]  - \beta
= 
\begin{bmatrix}
	(X_1' X_1)^{-1}(\underbrace{- X_1' (I - \Lambda_x) X_1 \beta_1}_{\Delta_{1}} + \underbrace{X_2' \Lambda_x X_1 \beta_2 }_{\Delta_{2}})\\
	(X_2' X_2)^{-1}( \underbrace{- X_2' \Lambda_x X_2 \beta_2}_{\Delta_{1}}+ \underbrace{X_2' (I - \Lambda_x) X_1 \beta_1}_{\Delta_{2}} )
\end{bmatrix}
$$
Activity bias now disappears because the researcher no longer stacks the fragmented data. However, other bias components remain. In particular, bias caused by cross-device substitution may still manifest itself in the omitted-variable-bias term ($\Delta_2$). Moreover, unlike in the common-effect model, here \textit{SIE} no longer guarantees attenuation bias. This is because the bias terms in different devices do not cancel out with each other when data from different devices enter separate models.

\subsubsection*{Extension to J Fragments.}\label{Jdevice}

Generalizing our discussion to the case where identities are split into $J\geq2$ fragments is straightforward. For the common-effect model, the bias is
\begin{equation}
	%\resizebox{0.93\hsize}{!}{$
	E[\widehat{\beta} | X_1, ..., X_J] - \beta = 
	\vartheta  \left\{
	\left[
	\sum_{j=1}^J  X_j' (\Lambda_j - \frac{1}{J}I)\eta 
	\right] \alpha
	+ 
	\left[
	\sum_{j=1}^J X_j' (\Lambda_j - I) X_j + 
	\sum_{j_1 \neq  j_2} X_{j1}' (\Lambda_{j1} + \Lambda_{j2}) X_{j2}
	\right] \beta
	\right\};
	%$}
\end{equation}
where $\vartheta$ is the $J \times J$ lower block diagonal of $(\widetilde{X}'\widetilde{X})^{-1}$. Under a more general set of symmetric treatment conditions: $\left\{ E[x_{j}]=E[x_{j'}],E[x_{j}^{2}]=E[x_{j'}^{2}],x_{j}\perp x_{j'},\forall j\neq j'\right\} $, we can show that $E[\beta|x_{1},...x_{J}]=\frac{\beta}{J}$. %, i.e., attenuation increases with the number of fragments. 

We can also extend the results corresponding to the device-specific-effect model to obtain, for the stacked-data version: 
\begin{equation}
	E[\widehat{\beta}|X_{1},...,X_{J}]-\beta=\vartheta\begin{bmatrix}X_{1}'(\Lambda_{1}-\frac{1}{J}I)\eta\alpha+X_{1}'(\Lambda_{1}-I)X_{1}\beta_{1}+\sum_{j\neq1}X_{1}'\Lambda_{1}X_{j}\beta_{j}\\
		\colon\\
		X_{J}'(\Lambda_{j}-\frac{1}{J}I)\eta\alpha+X_{J}'(\Lambda_{J}-I)X_{J}\beta_{J}+\sum_{j\neq J}X_{J}'\Lambda_{J}X_{j}\beta_{j}
	\end{bmatrix};
\end{equation}
and for the split-sample version:
\begin{equation}
	E[\widehat{\beta} | X_1, ..., X_J] - \beta= 
	\begin{bmatrix}
		(X_1' X_1)^{-1}
		(X_1' (\Lambda_1 - I) X_1 \beta_1 + 
		\sum_{j \neq1} X_1' \Lambda_1 X_j \beta_j )\\
		\colon \\
		(X_J' X_J)^{-1}
		(X_J' (\Lambda_J - I) X_J \beta_J + 
		\sum_{j \neq J} X_J' \Lambda_J X_j \beta_j)
	\end{bmatrix}.
\end{equation}
Results associated with the bias are analogous to the two-device version.

\subsection{Other Extensions}

While associated results are not reported, we have considered settings where the researcher is interested in other model forms, including treatment-effect type estimators and nonlinear models (e.g., logit). We have also considered cases where data may exhibit false matching, partial matching, or probabilistic matching.\footnote{
	Many of these analysis may be added to the paper in subsequent versions. They are available from the authors upon request.
} 
Our general finding across these models, settings, and specifications is that (a) the identity fragmentation bias continues to exists; (b) the bias structure is driven by some combination of the three components aforementioned; (c) in most cases the sign of the bias cannot be easily determined.

\section{Solution Comparison\label{solution}}

In this section, we compare three solutions to identity fragmentation bias: identity linking, experiment-based estimator adjustment, and stratified aggregation. Below we discuss their respective strengths and limitations.

	\subsection{Identity Linking}

Identity linking is often the go-to solution for practitioners. This approach aims to construct individual-level data from the fragmented ones, either by matching fragments that share the same characteristics (probabilistic matching) or by exchanging credentials associated with the same user (deterministic matching).

It is not hard to see the intuitive appeal of this approach. However, as the advertising industry responds to privacy concerns, many previously prevalent linking methods will become obsolete. For example, many probabilistic matching methods rely on third-party cookies, which will no longer work once Chrome stops their third-party cookie support.\footnote{
	https://headerbidding.co/universal-id-adtech/
} Other methods that rely on first-party data requires consent from consumers for data to be shared. As a result, these methods can only link part of the fragmented records in practice.

In some cases, partially linked data can exacerbate the fragmentation bias. Suppose the pooled data contain a proportion $r$ of fragmented users. The resulting estimator is the weighted average of
the ``pure'' estimators \citep{durbin1953note,theil1961pure}: 
$$
\widehat{\theta}^m=\omega\widehat{\theta}^{f}+(I-\omega)\widehat{\theta}^{l},\quad\text{where}\quad\omega=(r\widetilde{X}'\widetilde{X}+(1-r)X'X)^{-1}r\widetilde{X}'\widetilde{X},
$$
where $\widehat{\theta}^{f}$ is the estimator using only the fragmented data, while $\widehat{\theta}^{l}$ is the estimator obtained from the unfragmented data alone. It follows that $E[\widehat{\theta}^m|X]-\theta=\omega(E[\widehat{\theta}^{f}|X]-\theta)$. The problems with this estimator are that since $\omega$ is
a matrix, (i) $\omega$ does not necessarily increase ``monotonically'' with $r$; (ii) $\omega$ redistributes the bias in each element inside the vector $E[\widehat{\theta}^{f}|X]-\theta$. In simulations, we see that with intermediate ranges of $r$, $\widehat{\theta}^m$ is often further away from $\theta$, and sometimes the bias takes the reversed sign compared to $\widehat{\theta}^{f}$. The potential sign reversal and the amplified bias magnitude makes it even harder to bound the bias term.

\subsection{Experiment-Based Estimator Adjustment}

In Section \ref{randomization} and \ref{Jdevice}, we show that when \textit{SIE} holds, the slope estimator in a common-effect model is attenuated by the number of fragments per user. Thus, we can construct an unbiased estimator, $J\widehat{\beta}$, by multiplying the original estimator with the number of fragments. The confidence interval of this adjusted estimator is at least $\sqrt{J}$ times the confidence interval of the estimator with unfragmented data. This lack of power is caused by the loss of variation created by fragmentation.

To use this debiasing approach, a key question is when \textit{SIE} holds. The first condition in \textit{SIE} can be satisfied by running an experiment that completely randomizes treatments (e.g.,\ ad exposures) across devices. However, due to activity bias, the second condition is likely to be violated when the researcher estimates the actual treatment effect. Thus, we recommend researchers focus on intent-to-treat: in this case, both conditions in \textit{SIE} hold as long as a completely randomized experiment takes place. We also note that when users each has a different number of fragments, the estimator will assign more weights to users with more fragments. In practice, these are likely users with more activities.

\subsection{Stratified Aggregation}

A third solution is to aggregate fragmented units into higher-level groups, such that all fragmented identities of a consumer are collapsed within the aggregation. A simple form of aggregation is analyzing data at the geographic level, such as county, city, or stores \citep{, joo2013television, kalyanam2018, blake2018}. While simple aggregation provides robustness, it significantly reduces statistical power and restricts the ability to examine heterogeneity.

To alleviate the power problem, we propose stratified-aggregation. This approach uses observable user characteristic combinations to construct more granular bins, such that all fragments of a given user still fall into the same bin. For example, we can define bins by collapsing fragments that have the same zipcode $\times$ gender $\times$ age $\times$ education combination. The idea of using combinations of non-unique user characteristics to construct unique values for matching is similar to the technique used in probabilistic matching. However, unlike probabilistic matching, stratified aggregation does not require that matching be conducted at the individual level. In doing so, it allows for a more complete match. Given the caveats of partially matched data, we believe that stratified aggregation is more desirable. 

The granularity (and in turn, the effective sample size and power) increases with the number of observable user characteristics. For example, \citet{rocher2019} show that for the US population, a group constructed from 3 demographic variables consists of a unique individual 58\% of the time; with a 4th variable, the probability of the group consisting of one individual increases to 80\%. Thus, stratified aggregation is more effective when there are many observable user characteristics (columns) and when the number of observations is large (rows). We caution the readers that stratified aggregation may not perform as well when the covariates used for matching are measured with error (e.g.,\ Neumann et al. \citeyear{neumann2019}). 

As with experiment-based adjustment, stratified aggregation gives different weights to different users. In this case, users in smaller bins are weighted more. These users are likely consumers with more frequent activities or with rarer attributes.

\textbf{Summary.} Although identity linking is popular among practitioners, results based on partially linked data can be misleading. We recommend that research that uses identity linking services obtain metrics for the match rate, and interpret their estimates with caution. Experiment-based adjustment requires an experiment that completely randomizes treatment across fragment types (or exogenous variation that achieves the same effect), and only delivers a valid intent-to-treat effect under a common-effect model. Nevertheless, it is simple to implement when all the conditions hold. Stratified aggregation requires the least assumptions, and can be applied when the model of interest is nonlinear or even structural. However, it comes at the cost of statistical power.

\section{Empirical Application\label{application}}

The goal of this application is to show the magnitude of identity fragmentation bias in a realistic setting and to demonstrate the performance of stratified aggregation. We first describe the institutional background, and present data patterns that indicate the existence of identity fragmentation bias, then proceed to the estimation result discussion.

\subsection{The Context}

We focus on digital \textit{attribution}, namely, the problem of assigning credit for a positive outcome of interest (e.g., visits, engagements, conversions) across a variety of marketing and advertising touch-points \citep{li2014attributing,barajas2016,berman2018}. Any solution to the attribution problem relies on our ability to estimate the effect these marketing interventions have on the relevant outcomes. One can view this estimation as a digital analog of the classic marketing mix model, albeit on a much more granular scale. These estimators are calibrated on data obtained either via experimentation (e.g., Barajas et al. \citeyear{barajas2016}) or, more commonly, via observation. In either case, the unit of analysis is often the \emph{cookie}. Problems created by identity fragmentation are relevant in this context.

The data comes from an online seller of durable products. The goal of the exercise is to estimate the impact that various forms of online advertising have on customer engagement, measured by qualified visits to the firm's website. Our data has 391,195 observations, including the following variables: engagement ($Y$); display ads, search ads, and social media ads (X). 

The original data is a matched dataset, meaning that observations are at the individual consumer level. The data also records the devices where engagements and advertising exposures take place. These devices are classified as desktop, mobile phone, and tablets. We construct the fragmented data by dividing each observation into three fragments, each corresponding to a specific device. This data is observational and consequently we warn the reader from interpreting effects as causal. To the extent that marketing may be targeted, we conjecture that the implied targeting bias will be positive.

\subsection{Data Descriptives}

In total there are three types of bias components that contribute to the identity fragmentation bias: attenuation caused by measurement error on $Y$, omitted variable bias via fragmentation of $X$, and spurious correlation generated by activity bias and/or device substitution during purchases. The bias induced by measurement error on $Y$ always exists in fragmented data. Omitted variable biases will exist if $X_{j}$'s are correlated with each other. In our data, advertising exposures on different devices are not significantly correlated with each other, so omitted variable bias is likely to be a secondary issue. Therefore, we focus our attention on examining the evidence of activity bias and device substitution effects.

\begin{figure}[h!]
	\centering
	\caption{Ad Exposure across Devices\label{dist-x}}	
	\includegraphics[width=0.6\linewidth]{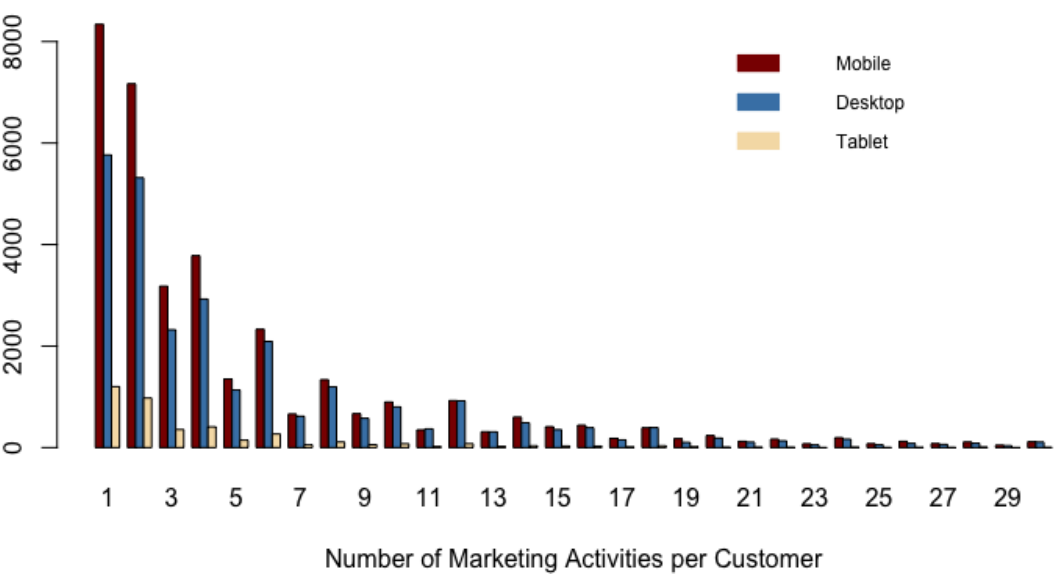}
\end{figure}

\begin{figure}[h!]
	\centering
	\caption{Engagement Distribution across Devices\label{dist-y}}	
	\includegraphics[width=0.6\linewidth]{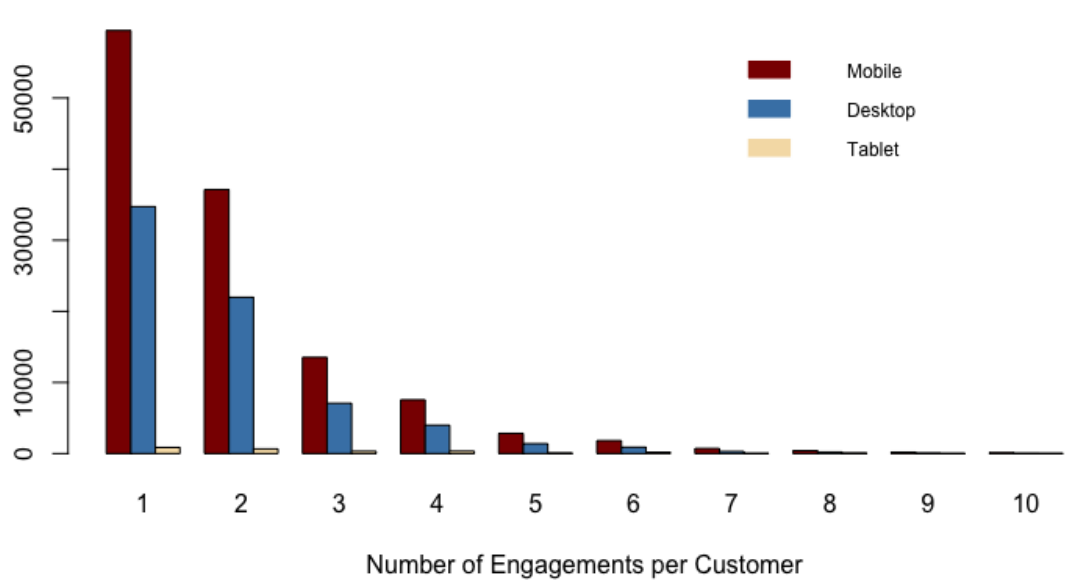}
\end{figure}

Activity bias exists if the average tendency of engagement on each device is positively correlated with ads on each device. Figure \ref{dist-x} shows the distribution of ads on different devices: it is heavy on mobile phones, less heavy on desktops, and minimal for tablets. The distribution of engagements (Figure \ref{dist-y}) shows a similar pattern across devices. The resemblance of the two histograms suggests the existence of activity bias.


Device substitution during purchases in response to advertising exposure will also contribute to the bias. Examining the cross-correlation matrix between $X_{j}$'s and $Y_{j}$'s is a simple approach to check if such substitution exists. Formally speaking, if there is no device substitution, that is, $s_{j}\perp(X_{1},...X_{J})$, then $Corr(Y_{j},X_{j})=\Lambda_{j}\cdot Corr(Y,X_{j'}),\forall j,j'$. This means that in the case of $J=3$, 
\[
\begin{aligned} & Corr(Y_{1},X_{1}):Corr(Y_{2},X_{1}):Corr(Y_{3},X_{1})=Corr(Y_{1},X_{2}):Corr(Y_{2},X_{2}):Corr(Y_{3},X_{2})\\
= & Corr(Y_{1},X_{3}):Corr(Y_{2},X_{3}):Corr(Y_{3},X_{3})=\Lambda_{1}:\Lambda_{2}:\Lambda_{3}.
\end{aligned}
\]
In other words, if device substitution is absent, we should expect different columns of the correlation matrix to be proportional to each other (same for rows). Table \ref{correlation matrix} shows that this is not the case.

A slightly more informal route to discerning device substitution involves comparing the diagonal of the matrix (roughly corresponding to the estimate using fragmented data) with the column-sums (corresponding to common-effect estimates using matched data). A diagonal element larger than the column-sum indicates that device substitution bias dominates the measurement error effect on $Y$. In Table \ref{correlation matrix}, we see that the diagonal elements are much larger than the column-sum in most cases. This strong device substitution pattern is likely to bias the estimates upward.

\begin{table}[h]
	\centering 
	\caption{Correlations Between Purchase Patterns and Advertising Exposure across Devices\label{correlation matrix}}
	
	\subcaption{Search}
	
	\centering 
	\begin{tabular}{c@{\hspace{3.6ex}}r@{\hspace{3.6ex}}r@{\hspace{3.6ex}}r}  
	\toprule
		& Search$_m$ & Search$_d$ & Search$_t$ \\
	 \midrule 		
		Y$_m$ & 0.2057 & -0.0439 & -0.0128 \\    
		Y$_d$ & -0.0446 & 0.1914 & -0.0172 \\    
		Y$_t$ & -0.0201 & -0.0117 & 0.2422 \\ 
	 \bottomrule
	\end{tabular}
	
	\vspace{0.2cm}
	\subcaption{Social}
	\centering 
	\begin{tabular}{c@{\hspace{3.8ex}}r@{\hspace{3.8ex}}r@{\hspace{3.8ex}}r}  
	\toprule
	& Social$_m$ & Social$_d$ & Social$_t$ \\    
	 \midrule 
	 Y$_m$ & 0.0642 & -0.0229 & -0.0046 \\   
	 Y$_d$ & -0.0163 & 0.0949 & -0.0039 \\    
	 Y$_t$ & -0.0064 & -0.0079 & 0.0715 \\     
	 \bottomrule
 \end{tabular}

	\vspace{0.2cm}
	\subcaption{Display}
	\centering	
	\begin{tabular}{crrr}  
		 \toprule
		 & Display$_m$ & Display$_d$ & Display$_t$ \\    
		 \midrule
		 Y$_m$ & 0.1363 & -0.0756 & -0.0192 \\    
		 Y$_d$ & -0.0488 & 0.2147 & -0.0131 \\    
		 Y$_t$ & -0.0224 & -0.0218 & 0.1518 \\    
		 \bottomrule
	\end{tabular}
	
\end{table}

\subsection{Estimation Results}

In Table 3, we present the estimates related to the common-effect and device-specific-effect models. A few points are worth mentioning. First, across all specifications, identity fragmentation leads to inflated estimates. In the common-effect specification, the magnitude of inflation ranges from 40\% (search ad effect) to 88\% (display ad effect). Identity fragmentation also leads to narrower confidence intervals compared to the true estimates: this is caused by the fact that the number of ``observations'' is tripled in the fragmented data. The combination of inflated point estimates and shrunk confidence interval leads to zero overlap between the fragmented and true estimates. Similar upward biases are seen in the device-specific-effect model, with the inflation in estimates ranging from 28\% (Social on Tablet) to 91\% (Display on Mobile). Overall, there are significant upward biases that could have serious consequences for marketing investment decisions if ignored.

\begin{table}[h!]
	
	\caption{Fragmentation Bias Estimates}
	\subcaption{Common Effects Model}
	
	\centering \begin{tabular}{rrrrrr}   \hline  & True Est & True SE & Fragmented Est & Fragmented SE & Bias Ratio ([3]/[1]) \\    \hline 
		Intercept & 0.9540 & 0.0024 & 0.3008 & 0.0009 & 0.3153 \\    Search & 0.2584 & 0.0025 & 0.3626 & 0.0016 & 1.4032 \\    Social & 0.1439 & 0.0039 & 0.2157 & 0.0026 & 1.4991 \\    Display & 0.0428 & 0.0007 & 0.0806 & 0.0005 & 1.8816 \\   \hline \end{tabular}
	
	\bigskip{}
	
	\subcaption{Device-Specific Effects Model} 
	\centering 
	\begin{tabular}{rrrrrr}   \hline  & True Est & True SE & Fragmented Est & Fragmented SE & Bias Ratio ([3]/[1]) \\    \hline 
		Intercept & 0.9535 & 0.0024 & 0.3006 & 0.0009 & 3.1720 \\    Search.m & 0.2564 & 0.0031 & 0.3699 & 0.0020 & 1.4430 \\    Social.m & 0.1191 & 0.0047 & 0.1859 & 0.0031 & 1.5603 \\    Display.m & 0.0396 & 0.0010 & 0.0759 & 0.0006 & 1.9160 \\    Search.d & 0.2741 & 0.0052 & 0.3733 & 0.0034 & 1.3622 \\    Social.d & 0.1985 & 0.0075 & 0.2836 & 0.0049 & 1.4285 \\    Display.d & 0.0456 & 0.0010 & 0.0849 & 0.0007 & 1.8597 \\    Search.t & 0.2394 & 0.0076 & 0.2979 & 0.0050 & 1.2445 \\    Social.t & 0.2516 & 0.0280 & 0.3238 & 0.0183 & 1.2868 \\    Display.t & 0.0449 & 0.0038 & 0.0813 & 0.0025 & 1.8116 \\      \hline \end{tabular}
\end{table}

\subsection{Stratified Aggregation}

We now turn to the examination of a proposed solution for the fragmentation bias. Since the original dataset does not contain customer demographics or other variables to aggregate around, we simulate a set of demographic variables and use them to augment the original data. In particular, for each true identity, we assign an \textit{MSA} variable uniformly drawn from one of 48 districts; \textit{Age} drawn uniformly within the {[}18, 82{]} range; and \textit{Income} drawn from one of 10 discrete income buckets.

Figure 3 shows the results of the stratified aggregation exercise. Although the $95$\% confidence intervals for the aggregated estimates are significantly larger in most cases, they contain the confidence intervals of true estimates as well as the point estimates from the unfragmented data. The aggregated estimates for the tablet ad effects do not perform as well since both purchases and ad exposures on the tablet are scarce. Even so, the results are significantly superior to those obtained when the identity fragmentation issues are ignored.

\begin{figure}[h!]
	\centering 
	\caption{Estimator Comparison across Different Data}
	\subcaption{Common Effects Model\label{estimates}}
	\includegraphics[width=0.8\linewidth, height=0.4\textheight]{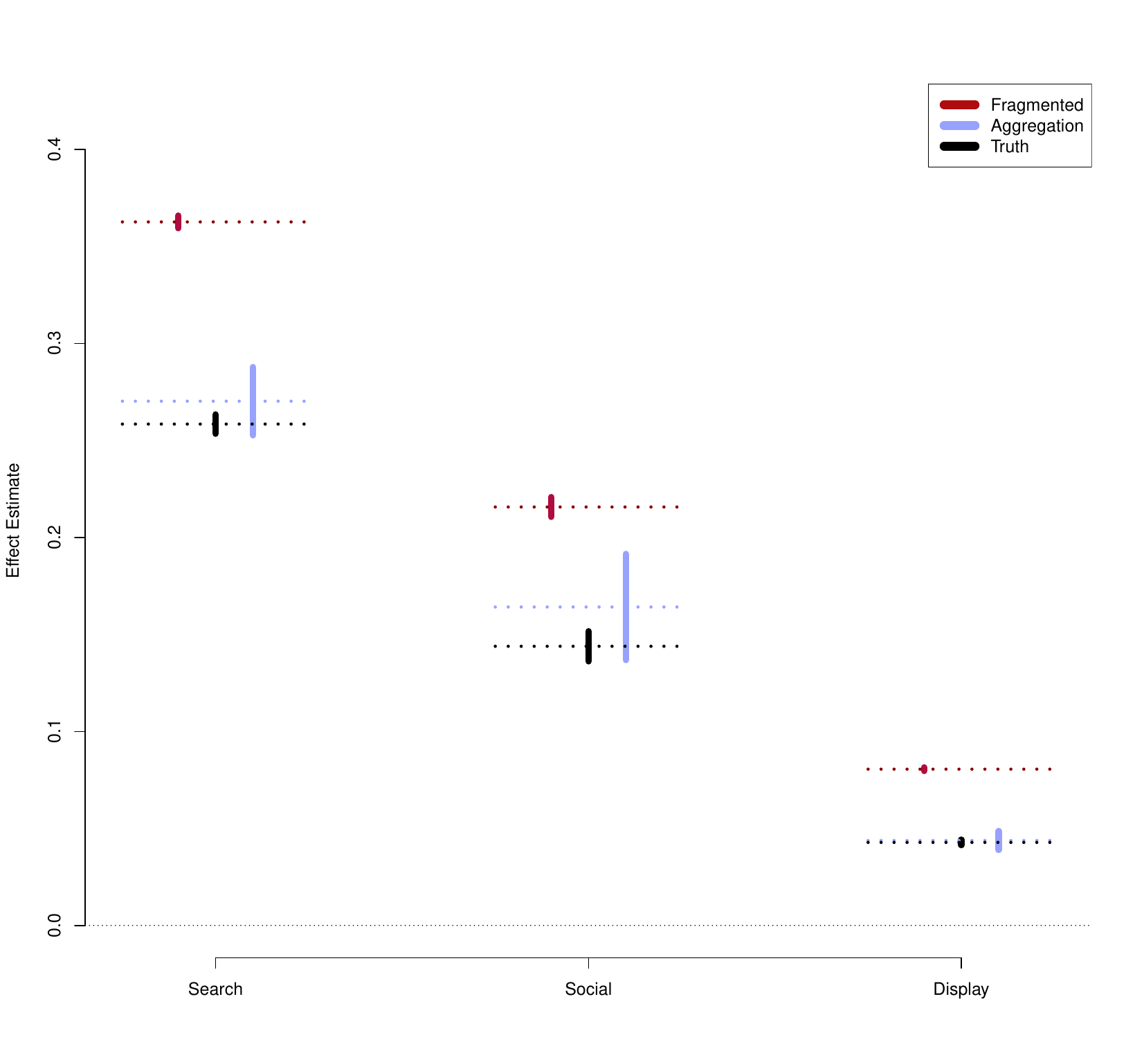}
	
	\subcaption{Device-Specific Effects Model} 
	
	\includegraphics[width=0.8\linewidth, height=0.4\textheight]{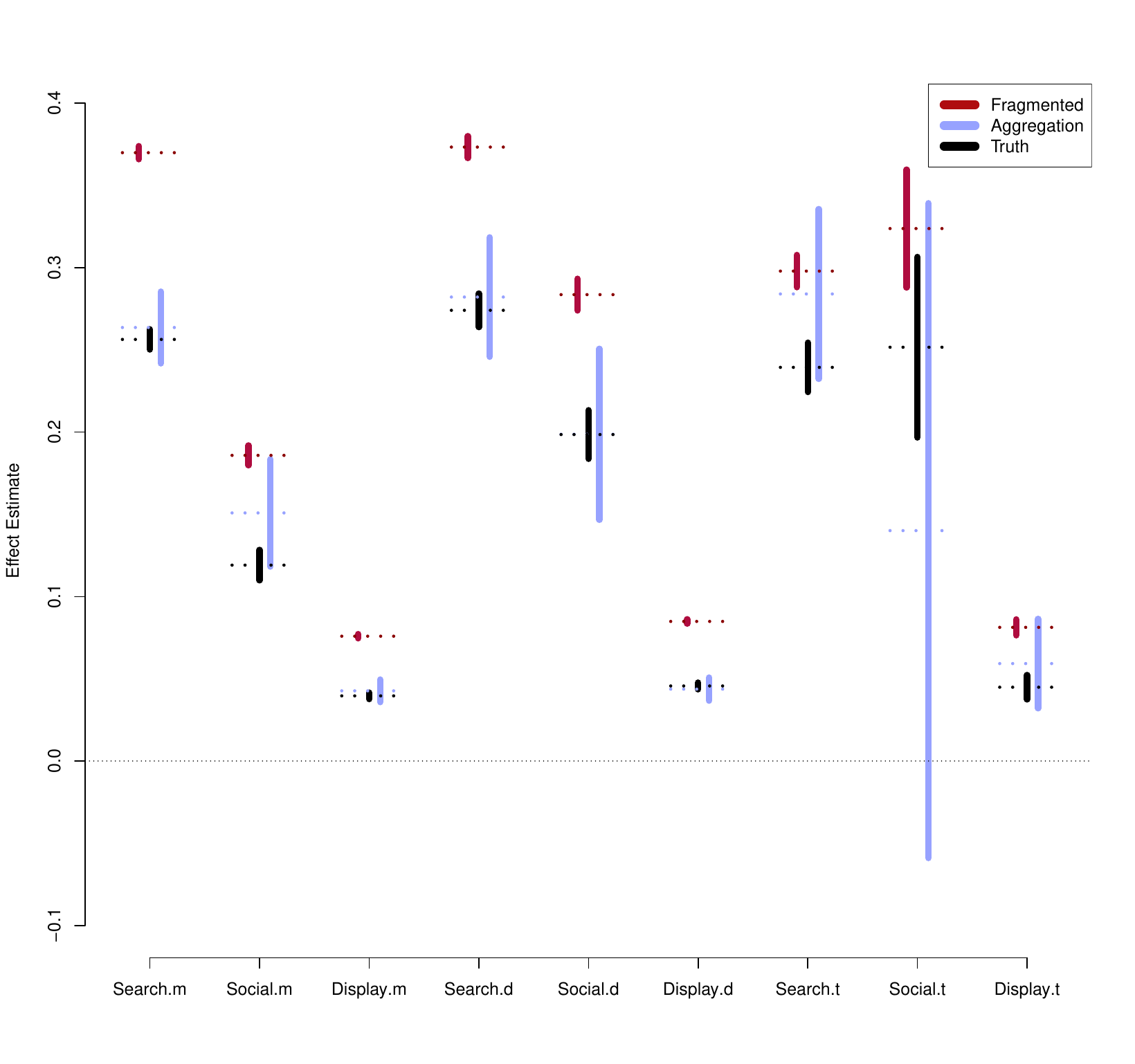}

	\floatfoot{\textit{Note:} The horizontal dashed lines indicate the estimated effect sizes, with black, red and blue representing the true (unfragmented), fragmented and stratefied aggregation based estimates. The vertical lines the 95\% confidence region.}
\end{figure}

\section{Discussion and Future Directions}

Using data with fragmented user identifiers leads to a persistent bias in estimators. We analytically characterize the bias and show that this bias can have an arbitrary sign. Based on this analytic characterization, we compare three debiasing solutions. Despite its popularity, identity linking can exacerbate the bias and complicate the effort of bounding the estimate when linking is incomplete. Experiment-based estimator adjustment relies on a set of strong assumptions and practicality requirements, but is simple and effective when these requirements are met. Stratified aggregation requires the least assumptions, and can be applied to obtain unbiased estimates for a more generalized set of models. However, it comes at a cost of statistical power.

Although we base our formal analysis on a linear-model setup, the results can also apply to generalized linear models, including discrete choice models and Poisson regressions. For example, one can rewrite a logistic regression as a linear regression of log odds on covariates. The attenuation bias is more likely to occur in discrete choice models, since estimated magnitudes are relative to the residual term. When the model is incorrectly specified, the residual term tends to be larger, and we expect the estimators to exhibit stronger attenuation. 

We look forward to future research that further investigates solutions to identity fragmentation bias. Without these solutions, the post-cookie world can render data analytics difficult or even impossible for most firms, while a few established firms who can obtain complete user data using a login wall can have a stronger incumbent advantage. By characterizing the bias and discussing existing solutions, we hope to contribute to this discussion and provide a more even playing field in the cookieless landscape.

\newpage

{
\singlespacing
\bibliographystyle{aea}
\bibliography{CrossDevice} 
}

\newpage
\begin{appendices}

\section{Proof for Identity Fragmentation Bias}
\setcounter{equation}{0} \renewcommand{\theequation}{A.\arabic{equation}}

\subsection{Common-Effect Model}
\label{proof-common}

Under the assumptions laid out in the paper, we have $\epsilon\perp(s,x_{1},x_{2})\Rightarrow E[g(x_{1},x_{2},s)\epsilon]=0,\forall g$. Therefore: 
\begin{equation}
	\begin{aligned}E[\widehat{\theta}|X_{1},X_{2}] & =(\widetilde{X}'\widetilde{X})^{-1}\cdot\left(\widetilde{X}'\cdot\begin{bmatrix}\Lambda_{x}\\
			I-\Lambda_{x}
		\end{bmatrix}W\cdot\widetilde{X}\right)\cdot\Omega\theta\\
		%& =(\widetilde{X}'\widetilde{X})^{-1}\cdot\left(\widetilde{X}'\cdot\begin{bmatrix}\Lambda_{x} & \Lambda_{x}\\
		%I-\Lambda_{x} & I-\Lambda_{x}
		%\end{bmatrix}\cdot\widetilde{X}\right)\cdot\Omega\theta\\
		& =\Omega\theta+(\widetilde{X}'\widetilde{X})^{-1}\cdot\left(\widetilde{X}'\cdot\begin{bmatrix}(\Lambda_{x}-I)I & \Lambda_{x}\\
			(I-\Lambda_{x})I & -\Lambda_{x}
		\end{bmatrix}\cdot\widetilde{X}\right)\cdot\Omega\theta.
	\end{aligned}
	\label{step1}
\end{equation}
Since $\widetilde{X}'\cdot\begin{bmatrix}(\Lambda_{x}-I)I & \Lambda_{x}\\
	(I-\Lambda_{x})I & -\Lambda_{x}
\end{bmatrix}
%=\begin{bmatrix}\eta' & \eta'\\
%X_{1}' & X_{2}'
%\end{bmatrix}\cdot\begin{bmatrix}I\\
%-I
%\end{bmatrix}\cdot\begin{bmatrix}\Lambda_{x}-I & \Lambda_{x}\end{bmatrix}
=\begin{bmatrix}0\\
	(X_{1}-X_{2})'
\end{bmatrix}\cdot\begin{bmatrix}\Lambda_{x}-I & \Lambda_{x}\end{bmatrix}$, equation (\ref{step1}) can be written as 
\begin{equation}
	\begin{aligned}E[\widehat{\theta}|X_{1},X_{2}] 
		& =\Omega\theta+(\widetilde{X}'\widetilde{X})^{-1}\cdot\left(\begin{bmatrix}0\\
			(X_{1}-X_{2})'
		\end{bmatrix}\cdot\begin{bmatrix}\Lambda_{x}-I & \Lambda_{x}\end{bmatrix}\cdot\begin{bmatrix}\eta & X_{1}\\
			\eta & X_{2}
		\end{bmatrix}\cdot\begin{bmatrix}\frac{1}{2}\alpha\\
			\beta
		\end{bmatrix}\right)\\
		%& =\Omega\theta+(\widetilde{X}'\widetilde{X})^{-1}\cdot\left(\begin{bmatrix}0\\
		%(X_{1}-X_{2})'
		%\end{bmatrix}\cdot\begin{bmatrix}(2\Lambda_{x}-I)\eta & (\Lambda_{x}-I)X_{1}+\Lambda_{x}X_{2}\end{bmatrix}\cdot\begin{bmatrix}\frac{1}{2}\alpha\\
		%\beta
		%\end{bmatrix}\right)\\
		& =\Omega\theta+(\widetilde{X}'\widetilde{X})^{-1}\cdot\left(\begin{bmatrix}0 & 0\\
			(X_{1}-X_{2})'(2\Lambda_{x}-I)\eta & (X_{1}-X_{2})'(\Lambda_{x}-I)X_{1}+\Lambda_{x}X_{2}
		\end{bmatrix}\cdot\begin{bmatrix}\frac{1}{2}\alpha\\
			\beta
		\end{bmatrix}\right)\\
		& =\Omega\theta+(\widetilde{X}'\widetilde{X})^{-1}\begin{bmatrix}0\\
			(X_{1}-X_{2})'(\Lambda_{x}-\frac{1}{2}I)\eta\alpha +(X_{1}'((\Lambda_{x}-I))X_{1}+X_{1}'X_{2}-X_{2}'\Lambda_{x}X_{2})\beta
		\end{bmatrix}.
	\end{aligned}
	\label{step2}
\end{equation}

\vspace{0.5cm}

To get $E[\beta|X_{1},X_{2}]-\beta$, only the bottom-right
block entry of $(\widetilde{X}'\widetilde{X})^{-1}$ needs to be calculated.
Let $\widetilde{X}'\widetilde{X}=\begin{bmatrix}2N & \eta'(X_{1}+X_{2})\\
	(X_{1}'+X_{2}')\eta & X_{1}'X_{1}+X_{2}'X_{2}
\end{bmatrix}\equiv\begin{bmatrix}A_{11} & A_{12}\\
	A_{21} & A_{22}
\end{bmatrix}$ and $(\widetilde{X}'\widetilde{X})^{-1}\equiv\begin{bmatrix}B_{11} & B_{12}\\
	B_{21} & \vartheta  
\end{bmatrix}$, then 
\begin{equation}
	\begin{aligned}E[\beta|X_{1},X_{2}] & =\beta + B_{21}\cdot0+\vartheta  \cdot\left((X_{1}-X_{2})'(\Lambda_{x}-\frac{1}{2}I)\eta\alpha+(X_{1}'(\Lambda_{x}-I)X_{1}+X_{1}'X_{2}-X_{2}'\Lambda_{x}X_{2})\beta\right)\\
		\Rightarrow E[\beta|X_{1},X_{2}] & -\beta =\vartheta  \cdot\left((X_{1}-X_{2})'(\Lambda_{x}-\frac{1}{2}I)\eta\alpha+(X_{1}'(\Lambda_{x}-I)X_{1}+X_{1}'X_{2}-X_{2}'\Lambda_{x}X_{2})\beta\right),
	\end{aligned}
	\label{step3}
\end{equation}
where 
\begin{equation}
	\begin{aligned}\vartheta  = & (A_{22}-A_{21}A_{11}^{-1}A_{12})^{-1}=\left[X_{1}'X_{1}+X_{2}'X_{2}-\frac{(X_{1}'+X_{2}')\eta\cdot\eta'(X_{1}+X_{2})}{2N}\right]^{-1}.\end{aligned}
\end{equation}
Note that $\vartheta^{-1}$ is always positive definite. Moreover, 
$$
\lim_{N \rightarrow \infty}\frac{1}{N} \vartheta^{-1} = Var [X_1] + Var [X_2] + \frac{1}{2} (E [X_1] - E [X_2])^2$$ 
is also positive definite.

\subsection{Device-Specific-Effect Model}
\label{proof-specific}

With a device-specific-effect model, we use the same notation to represent slightly different objects. The first change involves the covariates: $
\widetilde{X} \equiv 
\left[ \begin{array}{ccc}
	\eta & X_1 & \emptyset \\
	\eta&  \emptyset & X_2 
\end{array} \right].$ Second, the dimension of $\Omega$ changes from $k +1$ to $2 k + 1$; as before, the first element of $\Omega$ is 1/2 and the rest are 1. With the same notation, Equation (\ref{step1}) in Section \ref{proof-common} still holds. Now 
\begin{equation}
	\label{step1-sep}
	\begin{aligned}
		\widetilde{X}' \cdot 
		\begin{bmatrix}
			(\Lambda_x  - I) I & \Lambda_x  \\
			(I - \Lambda_x ) I &  - \Lambda_x 
		\end{bmatrix} 
		\cdot \widetilde{X}
		&= 
		\widetilde{X}' \cdot 
		\begin{bmatrix}
			I \\ -I
		\end{bmatrix} 
		\cdot 
		\begin{bmatrix}
			\Lambda_x  - I  & \Lambda_x 
		\end{bmatrix} 
		\cdot \widetilde{X}\\
		%&= 
		%\left(
		%\left[ \begin{array}{cc}
		%\eta' & \eta'  \\
		%X_1' & \emptyset \\
		%\emptyset & X_2' 
		%\end{array} \right]
		%\cdot 
		%\begin{bmatrix}
		%I \\ -I
		%\end{bmatrix} 
		%\right)
		%\cdot 
		%\left(
		%\begin{bmatrix}
		%\Lambda_x  - I  & \Lambda_x 
		%\end{bmatrix} 
		%\cdot 
		%\left[ \begin{array}{ccc}
		%\eta & X_1 & \emptyset \\
		%\eta&  \emptyset & X_2 
		%\end{array} \right]
		%\right)\\
		&=
		\begin{bmatrix}
			\emptyset & \emptyset & \emptyset \\
			X_1' (2 \Lambda_x  - I)\eta & X_1' (\Lambda_x  - I) X_1& X_1' \Lambda_x X_2 \\
			X_2' (I - 2 \Lambda_x)\eta & X_2' (I - \Lambda_x ) X_1& - X_2' \Lambda_x X_2 
		\end{bmatrix}.  
	\end{aligned}
\end{equation}
Combining Equation (\ref{step1-sep}) with Equation (\ref{step1}), 
$$
\begin{aligned}
	E [\widehat{\theta} | X_1, X_2] 
	= & 
	\Omega \theta + (\widetilde{X}' \widetilde{X})^{-1} \cdot 
	\begin{bmatrix}
		\emptyset & \emptyset & \emptyset \\
		X_1' (2 \Lambda_x  - I)\eta & X_1' (\Lambda_x  - I) X_1& X_1' \Lambda_x X_2 \\
		X_2' (I - 2 \Lambda_x)\eta & X_2' (I - \Lambda_x ) X_1& - X_2' \Lambda_x X_2 
	\end{bmatrix}  
	\cdot 
	\begin{bmatrix}
		\frac{1}{2} \alpha \\
		\beta_1 \\
		\beta_2
	\end{bmatrix} \\
	= & 
	\Omega \theta + (\widetilde{X}' \widetilde{X})^{-1} \cdot 
	\begin{bmatrix}
		0 \\
		X_1' (\Lambda_x  - \frac{1}{2} I)  \eta\alpha & X_1' (\Lambda_x  - I) X_1 \beta_1 & X_1' \Lambda_x X_2 \beta_2 \\
		X_2' (\frac{1}{2} I - \Lambda_x)  \eta\alpha  & X_2' (I - \Lambda_x ) X_1 \beta_1 & - X_2' \Lambda_x X_2 \beta_2
	\end{bmatrix}  
\end{aligned};
$$
thus 
$$
E [\widehat{\beta}] - \beta
= 
\vartheta   \cdot 
\begin{bmatrix}
	X_1' (\Lambda_x  - \frac{1}{2} I)  \eta\alpha & X_1' (\Lambda_x  - I) X_1 \beta_1 & X_1' \Lambda_x X_2 \beta_2 \\
	X_2' (\frac{1}{2} I - \Lambda_x)  \eta\alpha  & X_2' (I - \Lambda_x ) X_1 \beta_1 & - X_2' \Lambda_x X_2 \beta_2
\end{bmatrix}.
$$
Here as before, $\vartheta  $ is the bottom-right diagonal matrix of $(\widetilde{X}' \widetilde{X})^{-1}$; now it has dimension $2k \times 2k$. 
Note that 
$
\widetilde{X}' \widetilde{X} = 
\begin{bmatrix}
	2 N  & \eta' X_1 & \eta' X_1 \\
	X_1' \eta & X_1' X_1 & \emptyset \\
	X_2' \eta & \emptyset & X_2' X_2
\end{bmatrix}.
$
Let $A_{11} \equiv 2 N $, $A_{12} = \begin{bmatrix} \eta' X_1 & \eta' X_1 \end{bmatrix}$, 
$A_{21} = \begin{bmatrix} X_1' \eta \\ X_2' \eta \end{bmatrix}$, and 
$A_{22} = \begin{bmatrix} 
	X_1' X_1 & \emptyset \\
	\emptyset & X_2' X_2
\end{bmatrix}$. Then:
$$
\vartheta   
= (A_{22} - A_{21} A_{11}^{-1} A_{12})^{-1} 
=
\begin{bmatrix}
	X_1' (I - \frac{\eta \eta'}{2 N}) X_1 & - X_1' \cdot \frac{\eta \eta'}{2 N} \cdot X_2 \\
	- X_2' \cdot \frac{\eta \eta'}{2 N} \cdot X_1 & X_2' (I - \frac{\eta \eta'}{2 N}) X_2
\end{bmatrix}^{-1}.
$$

\end{appendices}

\end{document}